\documentstyle[12pt]{article}
\newcommand{\be}{\begin{equation}}
\newcommand{\ee}{\end{equation}}
\newcommand{\bea}{\begin{eqnarray}}
\newcommand{\eea}{\end{eqnarray}}
\newcommand{\bi}[1]{\bibitem{#1}}
\newcommand{\fr}[2]{\frac{#1}{#2}}

\begin{document}
\pagestyle{plain}
\pagenumbering{arabic}

\begin{center}
 {\large    \bf
            Reduction of the wave packet of a relativistic \\
              charged particle by emission of a photon
}

\vspace{0.5cm}

\normalsize
{ S.V. Faleev }\footnote{e-mail address:
S.V.Faleev@INP.NSK.SU}
{\it\footnotesize \\ Budker Institute of Nuclear Physics,
Novosibirsk, 630090, Russian Federation}
\vspace{0.5cm}

\end{center}
\begin{abstract}
{\footnotesize
The problem of reduction of the wave packet of a relativistic
charged particle
by emission of a photon is studied with help of the path integral approach.
A general expression for arbitrary order  correlation function of the
electromagnetic field is obtained. As a specific example
an ultrarelativistic electron circulating in a storage ring is considered.
It is shown that the longitudinal width of the electron wave packet
defined by characteristic difference of time between registrations of
two photons emitted by the electron
is of the order of the wave length of the photons.

}
\end{abstract}

\small
\section{ Introduction }\label{sec:1}

When detecting a photon emitted by the charged particle  we
measure the coordinate of the particle.
Following the general principles of quantum measurement theory it
results to the reduction of the wave packet of the particle.
However in our case the reduction of the wave packet occurs in the process
of emission of the photon and not in the process of measurement,
because the photodetector may be placed sufficiently far
from the particle path. The goal of the paper is to study which
way the reduction of the wave packet of the particle occurs and
to calculate the scale of the localization of the wave packet.

The technique used in the paper is the functional integral
formulation of relativistic quantum mechanics.  Within this
approach the interaction of the particle with quantum
electromagnetic field can be described with the
help of the "influence functional"
technique of Feynman and Vernon \cite{FV1,FV2}.

\section{Correlation function of electromagnetic field}

 We consider a relativistic charged particle moving in
external classical potentials and interacting with quantum
electromagnetic field. The state of the system is described by
a wave function which is the function of the coordinate of the
particle $q$ and the functional of variables of quantum
electromagnetic field
\be\label{1}
\Psi(q,t,\{A\})\equiv\langle q,A|\Psi\rangle \ .
\ee
The wave function is the projection of the state-vector of
the system $|\Psi\rangle $ onto the state of particle with
given coordinate $\langle q|$ and the state of electromagnetic field --
eigenstate of the field operator  $\hat{ A}(x)$ with the eigenvalue
$A(x) $:
\be\label{2}
\langle A|\hat{ A}(x)=\langle A|{ A}(x)\ .
\ee
 The field operator $\hat{A}(x)$ has the following standard expansion over the
operators of creation $a^+_{k\lambda}$ and annihilation $a_{k\lambda}$
of the photons with given momentum ${\bf k}$ and polarization $\lambda$
\be\label{3}
\hat{ A}(x)=\hat{ A}^-(x)+\hat{ A}^+(x)=\sum_{k\lambda}
(a_{k\lambda}e^{ikx} +  a^+_{k\lambda}e^{-ikx})\fr{{\varepsilon}_{k\lambda}}
{\sqrt{2\omega_kV}}\ ,
\ee
where $\hat{ A}^+(x)$ and $\hat{ A}^-(x)$ are the operators of
creation and annihilation of the photon at the point $x$,
${\bf \varepsilon}_{k\lambda}$ is the polarization vector,
$\omega_k=|k|$ -- the energy of the photon and $V$ -- the
normalization volume. The sum over momenta means the integration
\be\label{4}
\sum_k=\int \fr{d^3k V}{(2\pi)^3}\ .
\ee
The speed of light $c$ and Plank constant $\hbar$ in the paper
considered to be equal unity \\ $\hbar=c=1$.

 The time evolution of the wave function of the system is
controlled by the equation
\bea\label{5}
&i\partial_t\Psi(q,t,\{A\})= \\
&\{\sqrt{[-i{ \nabla}-e({
A}^{(ex)}(q,t)+{ A}(q))]^2+M^2}+eU^{(ex)}(q,t)\}\Psi(q,t,\{A\})
+\langle q,A|H_{em}|\Psi\rangle\ ,\nonumber
\eea
where $e$ and $M$ are the electrical charge and mass of the particle,
 $A^{(ex)}(q,t)$ and $U^{(ex)}(q,t)$ are the external vector and scalar
potentials, and
\be\label{6}
H_{em}=\sum_{k\lambda}\omega_k a^+_{k\lambda}a_{k\lambda}
\ee
is the hamiltonian of quantum electromagnetic field.
The external vector potential is given in  Coulomb gauge
\be\label{5.1}
\nabla A^{(ex)}=0\ .
\ee

 Let the initial wave function of the system be  factorized,
i.e. the particle be at pure state described by the wave
function $\psi_0(q)$ and electromagnetic field be at a state
$|\Phi_0\rangle$.
 The solution of the evolution equation (\ref{5}) can be
expressed in terms of the functional integral over paths of the particle:
\be\label{7}
\Psi(q_f,t,\{A_f\})=\int dq_i\psi_0(q_i) \int_{q_i}^{q_f} Dq e^{iS_0[q]}
\langle A_f| T\exp\{-i\int_0^t d\tau[H_{em}-e{\dot{q}}(\tau)
{\hat{A}}(q)]\} |\Phi_0\rangle\ .
\ee
Here $T$ is the symbol of chronological product. The action of relativistic
particle  $S_0[q]$ is given by the expression
\be\label{8}
S_0[q]=\int_0^t\!\!d\tau(-M\sqrt{1-\dot{q}^2}+
e\dot{q}A^{(ex)}(q,\tau)-eU^{(ex)}(q,\tau) ) \ .
\ee
The functional integrals in (\ref{7}) are evaluated over paths
$q(\tau)$ with endpoints $q(0)=q_i$ and $q(t)=q_f$.

We will consider only the case when characteristic energy of photons
emitted by the particle is much less than the energy of the particle
\be\label{8.2}
\omega_0\gamma^3\ll M\gamma\ ,
\ee
where $\omega_0$ is characteristic frequency of motion of
the particle in exernal fields and \\ $\gamma=(1-\dot{q}^2)^{-\fr{1}{2}}$ is
the relativistic factor. Otherwise one has to consider the quantum field
theory with arbitrary number of particle-antiparticle pairs.

The other restriction is that we are not allowed to consider
too short intervals of time. The integral over the paths
$q(\tau)$ in (\ref{7})
is defined as a product of the integrals over the coordinates $q(\tau_n)$
at discrete moments of time $\tau_n=dtn$, $n=1,2,3 ..$.
The functional integral (\ref{7}) is the solutions of the
equation (\ref{5}) only in the limit
\be\label{8.1}
 M\fr{dt}{\gamma}\gg 1\ .
\ee
Furthermore, if $dt M/\gamma \sim 1$ the functional integral is not
well-defined quantity. If the condition (\ref{8.1}) is satisfied
one has to evaluate the path integral (\ref{7}) by the steepest
descent method in semiclassical approximation by expanding the action
(\ref{8}) up to the quadratic term over the small deviation of
the path $q(\tau)$ from the classical, saddle path.

 On the other hand the time interval $dt$ is to be much less
than the characteristic time  $\sim(\omega_0\gamma)^{-1}$
of emission of the photon. The possibility to
choose the time interval $dt$ which satisfies both conditions
$\gamma/M\ll dt\ll (\omega_0\gamma)^{-1}$ is provided by the inequality
(\ref{8.2}).

 Let the initial state of quantum electromagnetic field be
the ground state $|0\rangle$. At each given path $q(\tau)$ the state
\be\label{9}
|\alpha(t,[q])\rangle\equiv T\exp\{-i\int_0^t d\tau[H_A-e{\dot{q}}(\tau)
{\hat{A}}(q)]\} |0\rangle
\ee
which arises in (\ref{7}) is a coherent state \cite{Glauber}.
The state is eigenstate of the annihilation operator ${\hat{A}}^-(x)$
\be\label{12}
{\hat{A}}^-(x)|\alpha(t,[q])\rangle={ A}^{(cl)}(x,t,[q])
|\alpha(t,[q])\rangle
\ee
with eigenvalue
\be\label{13}
{ A}^{(cl)}(x,t,[q])=i\int\fr{d^3k}{(2\pi)^3}\int_0^{t}d\tau
\fr{e}{2\omega_k}[{ \dot{q}}_{\tau}-{ n}(\dot{q}_{\tau} n)]
e^{ik(x-q_{\tau})-i\omega_k(t-\tau)}\ ,
\ee
where we denote ${\bf n}\equiv\fr{{\bf k}}{|k|}$ and for short
$q_{\tau}\equiv q(\tau)$.
As one can see, the field  $A^{(cl)}(x,t,[q])$ is the positive
frequency part of the classical vector potential  induced by the
motion of a point charge along the path $q(\tau)$ \cite{Landau2}.

We are interested in the two-photon correlation of radiation produced by the
ultrarelativistic particle. Let a photodetector be placed at the point $x_d$
on the tangent to the classical path of the particle.
 According to well-known results of quantum optics \cite{Glauber},
the probability $W(\Delta t)$ to detect two photons emitted by the
particle with time interval  between the registrations $\Delta t$
is proportional to the second order
correlation function of electromagnetic field
\be\label{14}
W(\Delta t)\propto \int dt_0
\langle
\hat{A}^+(x_d,t_0)\hat{A}^+(x_d,t_0+\Delta t)\hat{A}^-(x_d,t_0+\Delta t)
\hat{A}^-(x_d,t_0)\rangle \ .
\ee
Here we suppose that the time resolution of the photodetector is
much less than the characteristic time $\Delta t$ of changing of the
correlation function in r.h.s. (\ref{14}). The averaging
$\langle..\rangle$ in (\ref{14}) is taken over the
initial state of the system.
 The Heisenberg's creation $\hat{A}^+(x,t)$ and annihilation
$\hat{A}^-(x,t)$ operators are expressed through the Schr\"{o}dinger's ones
(\ref{3}) by standard way
\be\label{15}
{ \hat{A}}^{\pm}(x,t)=\{\tilde{T}e^{i\int_0^t d\tau H_{tot}(\tau)}\}
{ \hat{A}}^{\pm}(x)\{Te^{-i\int_0^t d\tau H_{tot}(\tau)}\}\ ,
\ee
where $\tilde{T}$ is the anti-chronological product symbol. The
total hamiltonian of the system $H_{tot}$ depends on time only if the
external potentials are time-dependent.

 Since the state $|\alpha(\tau,[q])\rangle$ (\ref{9}) is the
eigenstate of the operator ${\hat{A}}^-(x)$ one can easily
obtain the expression for the second order correlation function
of electromagnetic field
\bea\label{16}
&\langle
\hat{A}^+(x_d,t_0)\hat{A}^+(x_d,t_0+\Delta t)\hat{A}^-(x_d,t_0+\Delta t)
\hat{A}^-(x_d,t_0)\rangle = \\
&\int dq_f \int dq'_i\psi_0^*(q'_i) \int dq_i\psi_0(q_i)
\int_{q_i}^{q_f} Dq
\int_{q'_i}^{q_f} Dq'
 e^{iS_0[q]-iS_0[q']}F[q,q']\times \nonumber \\
& A^{*(cl)}(x_d,t_0,[q'])A^{*(cl)}(x_d,t_0+\Delta t,[q'])
A^{(cl)}(x_d,t_0+\Delta t,[q])A^{(cl)}(x_d,t_0,[q]) , \nonumber
\eea
where the influence functional is given by the following expression
\bea\label{17}
&F[q,q']\equiv\langle \alpha(t,[q'])|\alpha(t,[q])\rangle= \\
&\exp\left(
\int\fr{d^3k}{(2\pi)^3}\fr{e^2}{2\omega_k}\int_0^td\tau\int_0^tds
[\dot{q}_{\tau}
\dot{q}'_s-(n\dot{q}_{\tau})(n\dot{q}'_s)]
e^{-ik(q_{\tau}-q'_s)+i\omega_k(\tau-s)}\right)\times  \nonumber \\
& \exp\left(-
\int\fr{d^3k}{(2\pi)^3}\fr{e^2}{2\omega_k}\int_0^td\tau\int_0^{\tau}ds
\{[\dot{q}_{\tau}
\dot{q}_s-(n\dot{q}_{\tau})(n\dot{q}_s)]
e^{ik(q_{\tau}-q_s)-i\omega_k(\tau-s)}+ \right. \nonumber \\
& \left.[\dot{q}'_{\tau} \dot{q}'_s-(n\dot{q}'_{\tau})(n\dot{q}'_s)]
e^{-ik(q'_{\tau}-q'_s)+i\omega_k(\tau-s)}\}\right)\ .\nonumber
\eea
The upper limit of time $t$ in the path integrals (\ref{16}) is
an arbitrary time more than the largest argument of correlation function
$t_0+\Delta t$.

The expression (\ref{16}) can be evidently generalized to obtain the
correlation function of arbitrary order.

\section{           Reduction of the wave packet of
                  ultrarelativistic electron in storage ring
                        by synchrotron radiation}

Let us consider an ultralelativistic electron circulating in a storage ring
with frequency of revolution $\omega_0$
along an circular orbit of radius $R$. The  equilibrium circular path
$q_e(\tau)$ considered to be stable.

  Suppose that a point-like classical electron emits two photons
while passing through the sector of the orbit where radiation is
formed (i.e. when the angle between the velocity vector of the
particle and the vector "particle - photodetector" is little more than
$1/\gamma$). These two photons will be registrated by the detector
with the typical time interval between photocounts
being of the order $\Delta t\sim\lambda\!\!\bar{}\ /c$, where
$\lambda\!\!\bar{}\ \sim R/\gamma^3$ is the characteristic wave length of the
synchrotron radiation.
For a bunch of the point-like electrons with
the bunch length $\Delta l \gg \lambda\!\!\bar{}\ $
the time interval between photocounts
is defined by the width of the bunch $\Delta t\sim \Delta l/c$.
Thus, the characteristic scale $\Delta t$ of the two-photon correlation
function $W(\Delta t)$ (\ref{14}) gives us information about the longitudinal
width of the wave packet of single quantum electron.

 Our basic statement is that the longitudinal "size" of the electron
defined by characteristic difference of time between registrations of
two photons emitted by the particle is
of the order of the wave length $\lambda\!\!\bar{}\ $ of emitted photons.
It can be treated in the following way:
{\it the wave packet of the electron is reduced by the
emission of first photon  up to the wave length of the photon}.

Recently the two-photon function $W(\Delta t)$ of synchrotron radiation
was measured experimentally \cite{Vinokurov1}. It was obtain that the
characteristic width $\Delta t$ of the function coincides with
the resolution time of the measurement system which is much more than
$\lambda\!\!\bar{}\ /c$ but much less than the "natural bunch length".

 Now let us prove our statement. Consider the product of
classical fields which appears in the indegrand (\ref{16})
\be\label{18}
A^{(cl)}(x_d,t_0+\Delta t,[q])A^{(cl)}(x_d,t_0,[q]) \
\ee
at the time moment $t_0$ when the quantity $A^{(cl)}(x_d,t_0,[q])$
approaches its maximum. The basic point of our analysis is that
both  classical fields in the product (\ref{18}) depend on the same path
$q(\tau)$. If the path $q(\tau)$ is not too different
from the equilibrium circular path $q_e(\tau)$ the expression
(\ref{18}) would have a narrow peak at $\Delta t \sim \lambda\!\!\bar{}\ /c$
and vanish as $1/(\Delta t)^4$ at $\lambda\!\!\bar{}\ /c\ll \Delta t
\ll \omega_0^{-1}$. The same one can say about analogous to
(\ref{18}) product of classical fields in (\ref{16}) which
depends on the path $q'(\tau)$.

  Hence, to prove that the two-photon correlation function
$W(\Delta t)$ (\ref{14}) has a peak at $\Delta t\sim \lambda\!\!\bar{}\ /c$
we need to show that the dominating paths in the functional integrals
in (\ref{16}) are such, that the product of classical fields (\ref{18})
differs only slightly from that with $q=q_e$.
 To be specific, we have to show that characteristic  longitudinal
$\delta v_{\parallel}$ and transverse $\delta v_{\perp}$
deviation of the velocity $\dot{q} (\dot{q}')$ from the
equilibrium velocity $\dot{q}_e$ satisfy the conditions
\be\label{19}
(\delta v_{\parallel})^2\ll \fr{1}{\gamma^4}\ ;
\ \ \ \ \ (\delta v_{\perp})^2\ll\fr{1}{\gamma^2}\ .
\ee
 The conditions (\ref{19}) mean that fluctuations of the deviation angle
of velocity vector from the equilibrium velocity is much less
than $1/\gamma$ and fluctuations of the relativistic factor (and consequently
the energy) are small $\delta \gamma/\gamma\ll 1$.

 The functional integrals analogous to that in (\ref{16}) but without the
product of classical fields in the integrand were considered
by author in the work \cite{Faleev}. In semiclassical approximation,
when the difference of the paths $\delta q\equiv q-q'$ in the functional
integrals (\ref{16}) considered to be much less than the radius of curvature
of classical path (i.e. the radius of accelerator $R$),
the presence of the influence functional
(\ref{17}) in integrand (\ref{16}) leads to (infinite) renormalization
of the particle mass and appearance of the well-known \cite{Landau2} radiation
friction forces and so-called fluctuating forces in equation on the classical,
saddle path $\bar{q}_{cl}$, where $\bar{q}\equiv (q+q')/2$.
  The fluctuating forces, which accounts for the quantum discrete
character of radiation, have been originally introduced in the
works \cite{Sands,Lebedev} on statistical grounds.

 The fluctuating forces build up the synchratron oscillations of
the electron in accelerator. On the other hand, the forces of radiation
friction leads to damping of the oscillations. The uncertainty of the
energy $\delta \varepsilon\sim\gamma^2\sqrt{\omega_0M}$
\cite{Sands,Lebedev} associated with these oscillations
is much less than the energy of the electron
$\varepsilon=M\gamma$ provided by the condition (\ref{8.2}).
Hence, in the functional integrals (\ref{16}) the characteristic deviation
of the velocity of the classical path $\bar{q}_{cl}$ from the equilibrium
velocity  satisfies the requirement (\ref{19}).

Now, let us evaluate the quantum fluctuations of the velocity.
We calculate the functional integral over the paths $Dq (Dq')$
in (\ref{16}) by steepest descent method near the classical, saddle path
$\bar{q}_{cl}$. Each integration over the small deviation
$(q(\tau_n)-\bar{q}_{cl}(\tau_n))$ at time $\tau_n$ is performed
with the integrand which contains the multiplier
\be\label{20}
\exp\{-idtM[(\delta v_{\parallel})^2\gamma^3+(\delta
v_{\perp})^2\gamma]/2\} \ ,
\ee
arising from expansion of the kinetic part of the action (\ref{8})
over the deviation $\delta v\equiv \dot{q}-\dot{ \bar{q} }_{cl}$ up to
the quadratic term.
Because of the inequality (\ref{8.1}), the
corresponding projections $\delta v_{\parallel}$ and
$\delta v_{\perp}$ of typical quantum fluctuation of the velocity satisfy
the requirements (\ref{19}).

\section{Conclusion}

On the basis of the expression (\ref{16}) for the correlation
function of the electromagnetic field one can obtain the
two-photon $W(\Delta t)$ (or many-photon) correlation function  for
arbitrary system of external fields. For example, one can consider the
scattering of the sufficiently extended wave packet of the particle
on some potential. From the experience of previous section it
can be concluded that the "size" of the particle
defined by characteristic difference of time between registrations of
two photons emitted by the particle during the scattering process
is of the order of wave length of emitted photons.
One can say that
the wave packet of the particle is localized by the
emission of the photon  up to the wave length of the photon.

{\bf Acknowledgments}

The author is very grateful to I.V. Kolokolov and I.B. Khriplovich for
helpful discussions. Also I would like to thank N.A. Vinokurov and
T.V. Shaftan for useful conversations.
This work has been supported in part by the INTAS Grant 93-2492-ext.

\end{document}